\documentclass[11pt,twocolumn]{article}

\usepackage[dvips]{graphicx}
\usepackage{amssymb}
\newcommand{\eps}{\varepsilon}
\newcommand{\nn}{\nonumber}
\newcommand{\vphi}{\varphi}

\begin{document}

\title{Phase-separation in ion-containing mixtures in electric fields}

\author{
Yoav Tsori $^1$ Ludwik Leibler$^2$\\
$^1$Department of Chemical Engineering, Ben-Gurion University of the Negev, \\
P.O. Box 653, 84105 Beer-Sheva, Israel\\
$^2$Laboratoire Mati\`ere Molle \& Chimie (UMR 7167), ESPCI,\\
10 rue Vauquelin, 75231 Paris CEDEX 05, France}

\date{25 April, 2007}



\maketitle

\noindent {\bf When a liquid mixture is subjected to external electric fields, ionic
screening leads to field gradients. We point out that if the mixture is
initially in the homogeneous phase, this screening can bring about a
robust phase-separation transition with two main features: (i) the phase
separation is expected to occur in any electrode geometry, and (ii) the
voltage required is typically of the order of $1$ V and even less. We
discuss several applications of the effect relevant to the field of 
microfluidics, focusing on the
creation of a nanometer-scale lubrication layer in the phase-separation
process and the modification of the slip length.}

\section{Introduction}

The understanding and control of the phase behavior of liquid mixtures is
extremely important in everyday life, and is becoming equally important
in the field of microfluidics
\cite{micro_review1,micro_review2,micro_review3,micro_review4}. 
The behavior of
minuscule amounts of liquids has
drawn considerable attention lately, both from the aspect of basic
research as well as from the relevance to numerous applications utilizing
transport of small liquids drops \cite{weitz1}, mixing of
liquids \cite{ajdari1,ajdari2}, dielectrophoretic transport of colloidal
particles \cite{pohl}, etc.

As one confines himself to ever smaller regions of space, control over
the traditional parameters which govern the phase-behavior, such as
temperature, pressure, concentration and shear rate, becomes more and
more difficult. This control is essential in MicroElectroMechanical
Systems where the ultimate performance of a device is limited by the
lubrication of the surrounding liquid
\cite{klein1,klein2,granick1,granick2,granick3,isra1}.
Electric and magnetic fields, on the contrary, benefit
from size reduction since these fields are high near small conducting
objects, and therefore are excellent candidates for such a task.

Here, we describe a new type of phase-transition occurring in
ion-containing liquid mixtures under the influence of an external
electric field. It has been predicted long ago by Landau and Lifshitz
\cite{LL1} and later by Bedeaux, Mazur \cite{BM}, Onuki \cite{onuki} and
others, that a spatially uniform electric field can change the critical
temperature $T_c$ of mixture by a small amount, typically in the mK
range. In liquid mixtures containing dissociated ions, in contrast to the Landau case,
the electric field is screened, and the resulting gradients in the field and ion density
lead to strong electro-
and dielectrophoretic forces which tend to separate the mixture into its
components. The phase-transition is quite generic, and is virtually independent of
the electrode geometry. 

The model is presented below, and the resultant formulas for the
phase-separation derived. We further discuss the features of the effect
and its possible applications.

\section{Model}

Consider a binary mixture of two liquids A and B, with dielectric
constants $\eps_A$ and $\eps_B$, respectively, containing some amount of
dissociated positive and negative ions. When a voltage is applied on a
mixture which is initially homogeneous, there are two forces acting on the
liquid components. The
first one is a dielectrophoretic force: as the ions migrate towards the
electrode, the field is screened and therefore the high-$\eps$ liquid is
drawn to the electrodes. The second force is electrophoretic in nature -
the ions may have a chemical preference to one of the liquids, and, while
drifting to the electrodes, they will ``drag'' some liquid with them. The
two forces can work together or against each other. In general there is
also a process of recombination of positive and negative ions into a
neutral complex \cite{fisher1,fisher2,fisher3}, but in this simplified treatment this
process is not
allowed. We further restrict our attention to monovalent ions, each of
charge $1~e$.

We define $\phi$ as the relative A-liquid composition ($0<\phi<1$) and
$\rho^\pm$ as the number density of positive/negative ions.  We denote
$u_A^+$ and $u_B^+$ as the interaction energies of a positive ion with
the A and B liquids, respectively. The interaction energy between the
positive ion and the mixture is therefore
\begin{eqnarray}
u_A^+\rho^+({\bf r})\phi({\bf r})+u_B^+\rho^+({\bf
r})(1-\phi({\bf r}))=\nn\\
-\Delta u^+\rho^+\phi({\bf r})+const,\nn
\end{eqnarray} 
where $\Delta u^+\equiv u_B^+-u_A^+$ measures how much a positive ion prefers
to be in a A-liquid environment over a B-liquid one. Similar expression exists
for the interaction of the negative ions and the mixture. We can now write the
system free-energy as an integral $F=\int f[\phi,\psi]d^3r$, where on
the mean-field level the
free-energy density $f$ is given by
\begin{eqnarray}\label{FE}
f&=&f_b(\phi)-\frac12\eps(\phi)\left(\nabla\psi\right)^2
+\left(\rho^+-\rho^-\right)e\psi\nn\\
&+&k_BT\left[\rho^+\ln\left(v_0\rho^+\right)+
\rho^-\ln\left(v_0\rho^-\right)\right]-\lambda^+\rho^+\nn\\
&-&\lambda^-\rho^--\mu \phi-\left(\Delta 
u^+\rho^++\Delta 
u^-\rho^-\right)\phi\nn\\&+&const.
\end{eqnarray} 
In the above, $k_BT$ is the thermal energy, $\psi$ is the electrostatic
potential obeying the proper boundary conditions, $e$ is the electron
charge, $v_0$ is a molecular volume and $\lambda^\pm$ and $\mu$ are the Lagrange
multipliers 
(chemical potentials) of the positive and negative ions and liquid 
concentration, respectively. 
The mixture dielectric constant $\eps$ 
is assumed to depend on the composition through a quadratic constitutive
relation
$\eps(\phi)=\eps_c+\eps_1(\phi-\phi_c)+\frac12\eps_2(\phi-\phi_c)^2$, where
$\phi_c$ is the critical composition and $\eps_c$ is $\eps(\phi_c)$.
Finally, $f_b$ is the bulk energy density of
the mixture, which is taken here as a simple Landau expansion in the
deviation from the critical composition
\begin{eqnarray}\label{fb} 
\frac{v_0}{k_BT}f_b=\frac12 \frac{T-T_c}{T_c}(\phi-\phi_c)^2+
\frac{d}{24}(\phi-\phi_c)^4,
\end{eqnarray}
where $d$ is positive. This Landau energy has a
transition temperature $T_t$ given by $(T_t-T_c)/T_c=-\frac16
d(\phi-\phi_c)^2$. 

As can be seen from a systematic expansion of the free-energy in small
$\phi-\phi_c$ and examination of the {\it quadratic} term, at zero ionic
preference ($\Delta u=0$) and nearly uniform electric field $E_0$, the
transition temperature $T_t$ changes to $T^*$ by the Landau mechanism by
an amount  $T^*-T_t\sim v_0\eps_2E_0^2/k_B$. Similarly, with nonzero
preference $\Delta u$ and in the absence of field, $T_c$ changes by an
amount $T^*-T_t\sim (\Delta u/k_BT)^2\rho_0v_0$, where $\rho_0$ is the
bulk ion number density. While these shifts to the transition temperature
exist, they are negligible compared to the shift that we describe below
due to the dielectrophoretic and electrophoretic forces, which manifest
mathematically as {\it linear} terms in $\phi-\phi_c$ in the free-energy.

The free-energy expression Eq. {\bf \ref{FE}} depends on the four fields 
$\psi$, $\rho^\pm$, and the deviation from critical composition
$\vphi\equiv\phi-\phi_c$; the system equilibrium profile is given by the
variational principle with respect to these fields:
\begin{eqnarray}
\frac{\delta F}{\delta \vphi}&=&\frac{k_BT}{v_0}\left[\frac{T-T_c}{T_c}\vphi
+\frac16 d\vphi^3\right]-\frac{1}{2}(\eps_1+\eps_2\vphi)
\left(\nabla\psi\right)^2\nn\\
&-&\Delta u^+\rho^+-\Delta u ^-\rho^- 
-\mu=0\label{1stEL}\\
 \frac{\delta F}{\delta
\psi}&=&\nabla\left[\left(\eps_c+\eps_1\vphi+\frac12\eps_2\vphi^2
\right)
\nabla\psi\right]
\nn\\
&+&e\left(\rho^+-\rho^-\right)=0\label{2ndEL}\\
\frac{\delta F}{\delta \rho^\pm}&=&\pm
e\psi+k_BT\left(\ln\rho^\pm+1\right)-\Delta u^\pm\vphi
-\lambda^\pm=0.\label{3rdEL}
\end{eqnarray}
The second equation above is the Poisson equation.  For concreteness, we
consider two simple one-dimensional cases where the mixture is bounded by
either two walls at $x=0$ and $x=L$ with potentials $\psi(0)=V$ and
$\psi(L)=-V$, or bounded to the half space $x\geq0$ by a single wall at
$x=0$ with potentials $\psi(0)=V$ and $\psi(\infty)=0$. For simplicity,
we assume $\Delta u^+=\Delta u^-=\Delta u$, and that far enough from the
walls where the potential is zero, the system is coupled to a reservoir
at mixture composition $\phi_0$ and ion concentration $\rho_0$.  


~\\ \noindent {\bf Inadequacy of the linear Poisson-Boltzmann
approximation.} In field-induced phase-separation, the required fields
should be of the order $E\gtrsim 10^6$ V/$\mu$m \cite{TTL}. The
linearized Poisson-Boltzmann equation for a homogeneous mixture with
uniform dielectric constant $\eps$ gives the field $E$ near one wall at
$x=0$ with potential $V$ to be exponentially decaying:
$E=\lambda_D^{-1}Ve^{-x/\lambda_D}$, where $\lambda_D$ is the Debye
screening length given by $\lambda_D^{-2}=2\rho_0e^2/(\eps k_BT)$. For
typical values of $\lambda_D$ the field $E\propto V$ is thus too small
because the potential is small, $eV\ll k_BT$.
\begin{figure}
\centering
\includegraphics[scale=0.5,bb=80 180 515 580,clip]{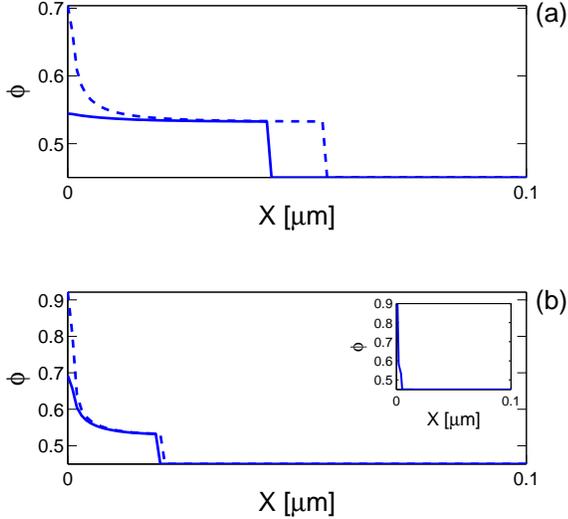}
\caption{{\footnotesize Composition profiles $\phi(x)$ for a mixture in the
vicinity of one wall at potential $V$ at $x=0$. Solid lines
correspond
to $V=0.2$V and dashed lines to $V=0.4$V. Far from the wall, the bulk
composition is $\phi=0.45$ and the bulk ion concentration corresponds to
pH 7 in (a) and pH 10 in (b). (Inset) $V=0.4$V and pH 12. In all plots the
temperature is $1$K above the transition temperature, the 
molecular volume is $v_0=8\times 10^{-27}$ m$^3$ and the dielectric
constants are $\eps_A=3$ and $\eps_B=2$.}} 
\end{figure}

As pointed above, the creation of high- and low-field regions due to
ionic screening leads to a dielectrophoretic force which tends to
``suck'' the high-$\eps$ material (assumed to be A) towards the region
with high field. If fields gradients are small, the mixtures composition
changes smoothly in the vicinity of the electrodes. However, if the field
gradients are large enough, the A-liquid composition crosses into the
unstable part of the phase-diagram, and a discontinuous composition
profile $\phi$ occurs, signifying a phase-transition \cite{TTL}. 
As a first approximation we can use the well-known analytical expressions
for the potential and ion distributions 
for a medium with uniform dielectric
constant $\eps=\eps_c$, and these can be substituted in Eq.
{\bf \ref{1stEL}}. Such an approximation is justified since
field gradients are mainly due to the ions and are much less influenced by
the mixture composition. 
As a result, analysis along classical lines \cite{LL2} predicts that the
transition temperature changes from $T_t$ to $T^*$ under the influence of
an external field, such that 
\begin{eqnarray}
\frac{T^*-T_t}{T_c}\simeq\left(\frac{|\eps_1|}{\eps_c}+ \frac{\Delta
u}{k_BT_c}\right)\frac{\rho_0v_0}{|\phi_0-\phi_c|}\nn\\
\times \exp\left(\frac{eV}{k_BT_c}\right)
\label{eq_DT}
\end{eqnarray}
This expression holds as long as $T^*$ is smaller than $T_c$; at all
temperatures $T>T_c$ the composition profile $\phi(x)$ varies smoothly
with no abrupt jump.

It is now clear that the dielectrophoretic force, proportional to the
dielectric mismatch $\eps_1$, and the electrophoretic force, proportional
to $\Delta u$, should be treated on equal footing. Note that for many
liquid pairs, $|\eps_1|/\eps_c\sim 1$ and $\Delta u/k_BT\sim 1$, and that
$\rho_0v_0$ is small: for a liquid with molecular volume $v_0=8\cdot
10^{-27}$ m$^3$ and ion content of pure water we have $\rho_0v_0\simeq
5\cdot 10^{-7}$. In addition, the denominator has a factor which measures
the distance from the critical composition, similar to the expression for
demixing in ion-free solutions. However, the most striking feature of Eq.
{\bf \ref{eq_DT}} is the exponential factor which can be huge -- already at
only $0.5$V and at room temperature $\exp(eV/k_BT)\simeq 4.8\cdot
10^{8}$. Different voltages change these figures dramatically, but clearly the
shift of the transition temperature can be very large. While the parameters determining
the Debye length all appear in Eq. {\bf \ref{eq_DT}}, $\lambda_D$ does not appear
explicitly due to the nonlinearity of the current theory.

Figure 1a shows the composition profile $\phi(x)$ calculated
numerically from Eqs. {\bf \ref{1stEL}}, {\bf \ref{2ndEL}} and {\bf \ref{3rdEL}}, 
for one wall at $x=0$ with potential $\psi(x=0)=V$, for two different
potentials  above the threshold for demixing: $V=0.2$V (solid line) and
$V=0.4$V (dashed line). The ionic content is the same as in a pH 7
solution (e.g., pure water). A clear front is seen separating A-rich (large
$\phi$) and A-poor (small $\phi$) domains. The A-liquid enrichment at the
wall is larger with the higher voltage. Figure 1b is the same, but the
ion density is much larger,  corresponding to pH 10. The phase-separation
front is created closer to the wall. Inset shows the profile when $V=0.4$V and
the pH is 12.

\section{Discussion}

From Eq. {\bf \ref{eq_DT}} we see that a liquid mixture phase-separates into
its components when put under the influence of an electric field in some
reasonable conditions. The dissociated ions in the solution 
are important because they bring
about large field gradients even in a
flat electrode geometry. Field gradients give rise to a dielectrophoretic
force which acts to pull the liquid with high dielectric constant towards
the region with high field (low dielectric component is attracted to the
low field). This tendency is accompanied by another equally important
electrophoretic tendency, where the ions attracted to the electrodes
preferentially ``drag'' with them one of the liquid components. This
second effect can enhance or negate the dielectrophoretic
phase-separation, depending on the solubility of the ions in the liquid
components.

For small enough potential, the composition of the A liquid component
(high dielectric constant) is enhanced close to the walls, but the
profile remains smooth. There exists a threshold voltage $V_c$ above
which phase-separation occurs, and the composition profile changes
dramatically - A-rich and A-poor domains are separated by a sharp
interface \cite{TTL}. The thickness of the A-rich domain can be extremely small,
and
depends nonlinearly on the ionic content in solution as well as on the
applied voltage.

For an ion-containing mixture, the nonlinear dependency on the voltage
means that increase in $V$ changes the field's spatial distribution in
addition to its amplitude. This is in contrast to ion-free mixtures,
where the applied voltage does not affect the field distribution, only
the amplitude \cite{TTL}. As a result, in ionic mixtures increase of the voltage
increases the composition difference between phase-separated domains and
may increase or decreases the thickness of the enrichment layer close to
the electrodes. Thus, the physics of the phase-separation considered here
is unique. 

The field-induced phase-separation has some important implications in
several circumstances. The first one relates to the rheological behavior
in systems with moving parts, that is {\bf field-controlled
lubrication}. This is reminiscent of pressure-induced melting in ice-skating, but
apparently richer. 
Consider two sub-micron-scale objects
sliding  past another so that the mixture confined between them is
sheared under conditions of low Reynold numbers
\cite{granick1,isra2,isra3,pgg,leger}. 
Let us denote the
viscosities of the A and B liquids by $\eta_{_A}$ and $\eta_{_B}$,
respectively. In the absence of field (mixed state) and under constant
applied external stress, the mixture will have the homogeneous viscosity
$\eta_m$, and the two surfaces will slide with a certain velocity $v_m$
with respect to each other. 

In the presence of electric field, (demixed state), the fluid exhibits
layers of different viscosities parallel to the walls
\cite{klein1,klein2,granick1}.
When the same
stress is applied across these layers, the surfaces move at a relative
velocity $v_d$. If the components' viscosities are very different,
$\eta_{_B}\gg\eta_{_A}$, the velocity gradient falls on a very thin layer of the less
viscous liquid, and it then follows that $v_d$ is much larger than $v_m$,
$v_d/v_m\approx \eta_{_B}/\eta_{_A}$. Essentially, the phase-separated mixture
has a smaller effective viscosity than the homogeneous one. This state is
reversible: when the field is turned off, the mixture becomes homogeneous again.
In a typical binary mixture of alkanes and siloxane oils (squalane
and polymethylphenylsiloxane), the viscosity ratio is about $10$,
thus the
effective viscosity of the demixed liquid is decreased by a factor $10$
as compared to the mixed solution. A different prominent example is a
water-glycerol mixture, where the velocity ratio is expected to be
$v_d/v_m\approx 1500$. Note though, that we do not expect a real
phase-transition
here but rather simply the creation of enrichment layers at the surfaces.
Other liquid pairs may prove to be more useful.
We also point out that the creation of viscosity layers at the
surface is equivalent to changing the slip length. Thus, in pressure-driven flows
and depending on the geometry one may be able to change, say, 
Poiseuille flow into plug flow, or vice versa, at a given moment and location.

Phase-separation could also be interesting in {\bf chemical reactions:} when
two or more chemical species
are undergoing a chemical reaction in a liquid environment, application
of an electric field can be used to phase-separate the liquids. This
can have two consequences: (i) If the reactant species exist
preferentially in one liquid component (say A), phase-separation will
lead to their accumulation into the A-rich environment, and to {\it
acceleration of reaction kinetics in a highly confined region of space}
($\lesssim 1\mu$m). (ii) If the reactant species prefer different liquid
components, after field-induced phase-separation, the reaction will be
{\it limited to the interface between coexisting phases and consequently slowed
down}. 

The phase-transition has some consequences in {\bf microfluidics optics}
\cite{micro_review4},
since in general the liquid components have different index of
refraction. Light wave will not be  deflected if it were to pass in a
homogeneous mixture and if the components are transparent enough.
However, once demixing occurs, interfaces between coexisting phases will
scatter, deflect or refract the light, and this could be used to create
optical switches or lenses in a microfluidic system coupled to an
external light source (ref. \cite{micro_review4} and unpublished data). 
Here again, the reversibility of the phase-separation is a boon.
Lastly, we mention that the electric field drops
off rapidly in the vicinity of highly charged objects in solutions, and
that the resulting field gradients could lead to local phase-separation
around charged colloids. For a colloid of size $R=1~\mu$m in ion-free
solution of dielectric constant $\eps=10\eps_0$, the field near the colloid's
surface is $E=Q/(\eps
R^2)$, and the charge $Q$ for separation is of the order of $1000$ $e$.
In salty solution with $\lambda_D\simeq 50$ nm, the field is $E\sim
V/\lambda_D$ and phase-separation is expected to occur when the colloid
potential is $V\lesssim 0.1$V.

This peculiar phase-separation could be further explored in the
directions outlined above. The dependence of demixing on the frequency of
applied external field, and the dynamics of field-induced phase-separation
should be studied as well.

\vspace{0.2cm}
\noindent Y. T. would like to thank P.-G. de Gennes for numerous
discussions on the subject, and for continuous support during his stay in
France. We benefited from many remarks and comments of M. E. Fisher, F. 
Tournilhac and B. Widom. We acknowledge useful discussions with D. Andelman, L.
Chai and J. Klein.
This research was supported by the Israel Science Foundation (ISF) under
grant no. 284/05.

\small{

}


\begin{thebibliography}{}


\bibitem{micro_review1} Squires TM, Quake SR (2005) {\it Rev Mod
Phys} 77:977--1026.

\bibitem{micro_review2} Gravesen P, Branebjerg J, Jensen OS (1993)
{\it J Micromech Microeng} 3:168--182.

\bibitem{micro_review3} Whiteside GM (2006)
{\it Nature} 442:368--373.

\bibitem{micro_review4} Psaltis D, Quake SR, Yang C (2006) 
{\it Nature} 442:381--386.

\bibitem{weitz1} Link DR, Anna SL, Weitz DA, Stone HA
(2004) {\it Phys Rev Lett} 92:054503-1--054503-4.

\bibitem{ajdari1} Stroock AD, Dertinger SKW, Ajdari A, Mezic I, 
Stone HA, Whitesides GM (2002) {\it Science}  295:647--651.

\bibitem{ajdari2} Joanicot M, Ajdari A (2005) {\it Science} 309:887--888.

\bibitem{pohl} Pohl HA (1978) {\it Dielectrophoresis}, (Cambridge Univ
Press, Cambridge, UK).

\bibitem{klein1} Raviv U, Laurat P, Klein J (2001) {\it Nature} 
413:51--54.

\bibitem{klein2} Raviv U, Klein J (2002) {\it Science} 
297:1540--1543.

\bibitem{granick1} Granick S, Lin Z, Bae S-C (2003) {\it Nature}
425:467--468.

\bibitem{granick2} Granick S, Lee H, Zhu Y (2003) 
{\it Nature Mat.} 2:221--227.

\bibitem{granick3} Van Alsten J, Granick S 
(1998) {\it Phys Rev Lett} 61:2570--2573. 

\bibitem{isra1} Bhushan B, Israelachvili JN, Landman U (1995)
{\it Nature} 374:607--616.


\bibitem{LL1} Landau LD, Lifshitz EM (1957) in {\it Elektrodinamika
Sploshnykh Sred} (Nauka, Moscow) Ch. II, Sect. 18, problem 1.

\bibitem{BM} Sengers JV, Bedeaux D, Mazur P, Greer SC (1980) 
{\it Physica A} 104:573–-594.

\bibitem{onuki} Onuki A (1995) {\it Europhys Lett} 29:611–-616.

\bibitem{fisher1} Levin Y, Fisher ME (1993) {\it Phys Rev Lett} 
71:3826--3829.

\bibitem{fisher2} Levin Y, Fisher ME (1996) {\it Physica A} 
225:164--220.

\bibitem{fisher3} Fisher ME (1994) {\it J Stat Phys} 75:1--36.

\bibitem{TTL} Tsori Y, Tournilhac F, Leibler L (2004) {\it
Nature} 430:544--547.

\bibitem{LL2} Landau LD, Lifshitz EM (1980) {\it Statistical Physics}
(Butterworth-Heinmann, New-York), 2nd Ed.

\bibitem{isra2} Israelachvili JN, Tabor D (1973) {\it Nat Phys
Sci} 241:148--149.

\bibitem{isra3} Horn RG, Israelachvili JN (1981) {\it J Chem
Phys} 75:1400--1411.

\bibitem{pgg} de Gennes, P-G (2002) {\it Langmuir} 18:3413--3414.

\bibitem{leger} Pit R, Hervet H, L\'{e}ger L (2000) {\it Phys Rev
Lett} 85:980--983.



\end{thebibliography}
\end{document}